\documentclass[a4paper,11pt]{article}
\usepackage{jinstpub} 
\usepackage[english]{babel}

\usepackage{subcaption}

\title{\boldmath Riptide: a proton-recoil track imaging detector for fast neutrons}

\author[a,b]{C.~Pisanti,}
\author[a]{A.~Berardi,}
 \author[c,b]{P.~Console Camprini,}
 \author[d]{F.~Giacomini,}
 \author[a,b,1]{C.~Massimi,\note{Corresponding author.}} \emailAdd{cristian.massimi@unibo.it}
 \author[b]{A.~Mengarelli,}
 \author[e,f]{A.~Musumarra,}
 \author[f]{M.~G.~Pellegriti,}
 \author[a,b]{R.~Ridolfi,}
 \author[b]{R.~Spighi,}
 \author[c]{N.~Terranova,}
 \author[a,b]{and M.~Villa}
  \affiliation[a]{Department of Physics and Astronomy, University of Bologna \\Via Irnerio 46, 40126 Bologna, Italy}
 \affiliation[b]{INFN-Bologna division,\\Via Irnerio 46, 40126 Bologna, Italy}
 \affiliation[c]{ENEA,\\via Enrico Fermi, 45 Frascati, Italy}
 \affiliation[d]{INFN-CNAF,\\Viale Berti Pichat, 6/2, Bologna, Italy}
 \affiliation[a]{Department of Physics and Astronomy, University of Catania \\Via Santa Sofia 64, I-95123 Catania, Italy }
 \affiliation[b]{INFN-Catania division,\\Via Santa Sofia 64, I-95123 Catania, Italy }

\abstract{Riptide is a detector concept aiming to track fast neutrons. It is based on neutron--proton elastic collisions inside a plastic scintillator, where the neutron momentum can be measured by imaging the scintillation light. More specifically, by stereoscopically imaging the recoil proton tracks, the proposed apparatus provides neutron spectrometry capability and enable the online analysis of the specific energy loss along the track. In principle, the spatial and topological event reconstruction enables particle discrimination, which is a crucial property for neutron detectors. In this contribution, we report the advances on the Riptide detector concept. In particular, we have developed a Geant4 optical simulation to demonstrate the possibility of reconstructing with sufficient
precision the tracks and the vertices of neutron interactions inside a plastic scintillator. To realistically
model the optics of the scintillation detector, mono-energetic protons were generated inside a $6\times6\times6$ cm$^3$
cubic BC-408 scintillator, and the produced optical photons were propagated and then recorded on a scoring plane corresponding to the surfaces of the cube. The photons were then transported through an optical system to a $2\times2$ cm$^2$ photo sensitive area with 1 Megapixel. Moreover, we have developed two different analysis procedures to reconstruct 3D tracks: one based on data fitting and one on Principal Component Analysis. The main results of this study will be presented with a particular focus on the role of the optical system and the attainable spatial and energy resolution.}


\begin{document}
\maketitle
\flushbottom
\section{Introduction}
\label{sec:intro}
Neutron detection systems have continuously improved in the past decades~\cite{PIETROPAOLO20201} and can be considered well-established radiation detectors. Advancements in neutron detectors are related to the instrument developments of particle detectors, sharing the same underlying cutting-edge technology. 
However, while tracking charged particle is a consolidated technique, neutron trackers are still being developed. For instance, only a few detector concepts or feasibility studies are present in the literature, e.g.~\cite{Hu2018,VALLE2017556,GIOSCIO2020162862,Wang2020,MILLER2003}. Tracking fast neutrons~\cite{WANG2013} requires a complete momentum reconstruction of the detected neutrons. In this respect, the advantage of using two-particle reactions is evident, and neutron--proton (n--p) elastic scattering represents the simplest interaction to exploit. 

State-of-the-art approaches for neutron tracking by using n--p single and double scattering have been proposed by the groups working on the SONTRAC~\cite{Mitchell2021,DENOLFO2023} and the MONDO~\cite{VALLE2017556,GIOSCIO2020162862,Marafini2017} projects. Both systems are based on multiple n--p scatterings: SONTRAC is being developed for Solar MeV-neutron detection in spacecrafts while MONDO aims to track fast neutrons produced in Particle Therapy treatments. In both cases, recoiling protons are detected using a matrix of plastic scintillating fibers. Then, the produced light is either amplified using a triple GEM-based image intensifier (alternatively by single photon avalanche diode arrays) in MONDO or read-out by silicon photomultipliers in SONTRAC. The challenging aspects of these pioneering projects are related to the read-out of channels and data transfer, as well as to the efficiency cut-off.

With the aim of addressing the challenge, a few years ago we proposed a novel detector concept named RIPTIDE (RecoIl Proton Track Imaging DEtector)~\cite{Musumarra_2021,Massimi_2022,Camprini_2023} . It consists of a monolithic plastic scintillator coupled to CMOS imaging devices. More in detail, the light output produced in the fast scintillator is used to perform a complete reconstruction in space and time of the n--p scattering event. In fact, by stereoscopically imaging the recoil-proton tracks the proposed apparatus can achieve neutron spectrometry capability, thus enabling real-time analysis of the specific energy loss along the charged particle track. In summary, Riptide is based on the same physics interaction exploited in SONTRAC and MONDO, but it is conceived to be simpler, less expensive and possibly featuring a more efficient detection geometry. 

In this contribution we report some feasibility studies -- based on Monte Carlo simulations -- showing the viability of the proposed idea. In particular, section~\ref{sec:PRTI} reports a detailed description of the proposed detector. Section~\ref{sec:Geant4} shows the Monte Carlo studies performed with Geant4~\cite{AGOSTINELLI2003} for the transport of optical photons, and section~\ref{sec:optic} illustrates the main characteristics of the optical system which collects the photons. Section~\ref{sec:tracks} describes the track reconstruction algorithms and show the results of preliminary analyses performed.
\section{Proton recoil technique and neutron tracking in Riptide}\label{sec:PRTI}
The basic tool for full neutron momentum reconstruction in Recoil Proton Track Imaging (RPTI) detectors is the two-body kinematics, where the neutron energy $E_n$ is related to the proton recoil angle and energy ($\theta_p$,$E_p$) by the relationship: \begin{equation}\label{eq:RPTI}
E_n=E_p/\cos^2(\theta_p).    
\end{equation}
To determine the neutron energy, and most importantly to retrieve its trajectory, eq.~(\ref{eq:RPTI}) can be used to reconstruct with good efficiency the neutron momentum from:
\begin{itemize}
    \item single n--p scattering, when the primary production point of the neutron trajectory is known
(e.g. point-like target in fixed-target experiments);
\item double or multiple n--p scattering in the detector active volume (general case).
\end{itemize}
Both cases are of interest for Riptide, as sketched in previous publications~\cite{Camprini_2023,Massimi_2022,Musumarra_2021}. The Riptide detector concept consists of a cubic (216 cm$^3$) plastic scintillator (BC-408/EJ-200~\cite{BC408}) surrounded by two (or more) optical systems of lenses focusing the proton tracks into CMOS cameras. 

From preliminary feasibility studies based on Geant4 simulations, we have estimated absolute detection efficiency with the full optics over the energy region of interest~\cite{Camprini_2023}, and studied the impact of carbon share in BC-408. Finally, other kinds of background, for instance due to gamma rays and charged particles, can be rejected by $dE/dx$ track characterization.

The next step in our detector study is the attempt to perform a 3D tracking. In this field, a few successful proof-of-principle experiments demonstrated the possibility of 3D tracking of charged particles by scintillation light~\cite{Filipenko14,YAMAMOTO2021}.
\section{Geant4 simulations and optical photon transport}\label{sec:Geant4}
The detection concept proposed in this feasibility study has been supported through Monte Carlo simulations based on the Geant4 toolkit~\cite{AGOSTINELLI2003}, version 10.4.2. The scintillation cube (of size $6\times6\times6$ cm$^3$) has been described with a composition of plastic polyvinyl toluene. The properties of the specific material have been considered representative of the BC-408 and have been retrieved from the NIST database~\cite{NIST} available among the code libraries: composition name is G4\_PLASTIC\_SC\_VINYLTOLUENE, with stoichiometric ratio C/H$=9/10$, density 1.032 g/cm$^3$ and ionization energy of 64.7 eV.
Transport simulations by Geant4 have been ruled by standard physics lists based on models. In particular, neutron--proton interaction has previously been tested within the framework of other experiments and it is considered as validated in this context and reliable in the current energy range (see for instance~\cite{TERRANOVA2020}).

Moreover, the optical photon production and transportation features of Geant4 have been enabled. In fact, suitable physics libraries are available to be used together with proper material characteristics to be defined. In accordance to the BC-408 data sheet~\cite{BC408}, a number of $10^4$ optical photons per MeV were produced along the proton track and transported inside the plastic scintillator. In addition, a refractive index of 1.59 has been considered as well as a very long absorption length, compared to the cube dimension. In this way, the photons can be originated along the tracks of charged particles -- mainly protons -- ionizing the scintillator. Then they are emitted uniformly into the $4\pi$ solid angle with a random linear polarization perpendicular to their momentum direction.

Geant4 simulations regarded primary particles as well as secondary photons. Source particles were either protons or neutrons.

Neutrons as source particles were used to simulate the scintillator response in terms of energy transfer from neutron to protons and the corresponding transport in the medium. Neutron--proton and neutron--carbon interactions were considered and registered~\cite{Camprini_2023}.

Here we report on protons as source particles. In fact, proton sources are useful to directly investigate the response of the scintillator to the proton-induced ionization. In particular, proton energy and direction were imposed: sets of mono-energetic protons ($5<E_p<100$ MeV) were randomly originated inside a cubic volume ($2\times2\times2$ cm$^3$) in the center of the plastic scintillator with isotropic momentum direction. Each source characteristic could be registered, together with the range in the medium.

Concerning secondary particle tracking, the propagation of protons in the plastic scintillator cube and the related energy deposition induced the production of optical photons. The position and the direction of the photons reaching the surface were
recorded for further analysis. In addition, each photon reaching the boundary was associate to its production point -- namely the position in which it was produced by ionization. Merging all production  points for the escaping photons could yield the shape of the ionization track related to the proton source. 

\section{Study of a simple optical system}\label{sec:optic}
\begin{figure}[h!]
\centering
\includegraphics[width=0.8\textwidth]{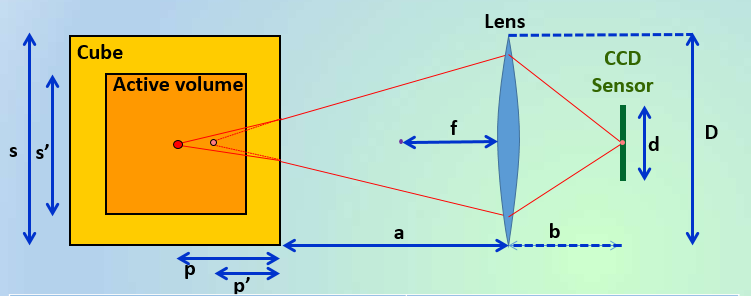}
\caption{Sketch of a simple setup for the simulation of photon transport. 
\label{fig:simple_camera}} 
\end{figure}

As a first step, we simulated the response of a simple optical system made of a lens and a large surface sensor to maximize the photon collection efficiency.
This simple case, sketched in fig. \ref{fig:simple_camera} allowed us to study the propagation of photons, including light refraction and aberrations. As mentioned above in section~\ref{sec:Geant4} in the Geant4 simulations, an inner active volume for the proton detection with side 40 mm (s' in fig.~\ref{fig:simple_camera})  was considered. Lens and sensor parameters were set in a way to cover the whole active volume. By considering a realistic sensor size of 20 mm (d in fig.~\ref{fig:simple_camera}), a magnification factor of 0.5 was chosen. 

The remaining parameters were calculated starting from these constraints and using the thin lens equation and Snell's law, the refraction index between the scintillator and air being $n = 1.59$ (see section~\ref{sec:Geant4}).  

The propagation of photons through this simple setup produces different types of
aberrations, the predominant one being the spherical aberration. To minimize it, the simulation was repeated while reducing the lens radius (D in fig.~\ref{fig:simple_camera}) from half the side of the scintillator cube to 5 mm. 

For each face of the scintillator cube, the simulation provided 2D projections at the sensor position (i.e. at a distance a+b in fig.~\ref{fig:simple_camera} from the outer surface of the BC-408 cube). These 2D images were then used as input to reconstruct the 3D proton tracks, and the results were compared to Monte Carlo data.
\section{Proton-track reconstruction}\label{sec:tracks}
Starting from the 2D projections of the 3D tracks described above, we have used two different approaches to reconstruct the original proton tracks. A first and simple method is based on linear interpolation of the 2D distributions as described in Section~\ref{sec:tracks1}. A second and different approach based on Principal Component Analysis (PCA) is described in Section~\ref{sec:tracks2}.

\subsection{Linear interpolation}\label{sec:tracks1}
It is well known that a distribution in two dimensions fitted to a straight line can have undesired sensitivity to outlying points, see ~\cite{NumericalRecipes} and references therein. In fact, as exceptions to a Gaussian model for experimental error, outlier points  can cause a wrong estimation of the central value. To prevent such a bias, we have used a robust statistical estimator which is insensitive to small departures from the idealized assumptions for which the estimator is optimized. In particular, we fitted the data minimizing the Least Absolute Deviations (LAD) instead of the Least Squares:
    \begin{equation}
        LAD = \sum| y_0 - y_p |
    \end{equation}

\begin{figure}[h!]
\centering
\includegraphics[width=\textwidth]{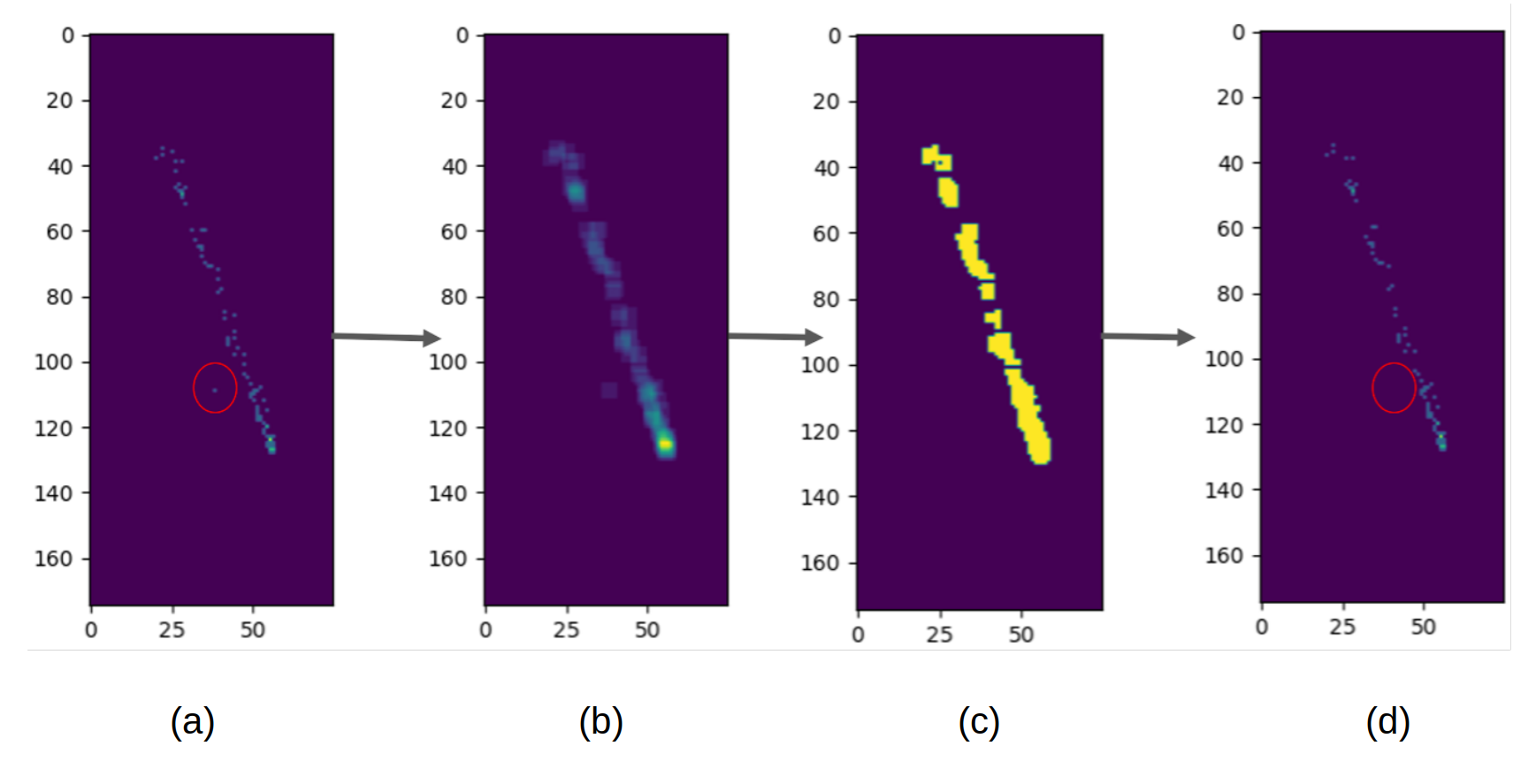}
\caption{An illustration of the spurious signal removal process (as the pixel highlighted by the red circle in the left panel) in a 2D projection of a proton track obtained from MC simulations using the simple optical system described above. 
From left to right, the 2D-projection as it appears on the sensor (a) is blurred with a mean filter to enhance the pixel density (b). A mask is created from the blurred image to identify the most dense region, which represents the effective region where the track projection is (c). By applying the mask into the initial image it is possible to remove the spurious pixel as shown in (d). 
\label{fig:outlayers}}
\end{figure}

LAD  is known to be less sensitive to non-Gaussian fluctuation. $y_0$ is the actual point and $y_p$ is the value predicted from the fit.
 Once the direction of the main projection was identified by the fit, its length was determined excluding the outliers, i.e.  the values sufficiently far from the fitted line. An example of an analysis procedure to remove spurious pixels in a reconstructed proton track is shown is fig.~\ref{fig:outlayers}. 
 
 From each 2D projection, the length of the projected segment was obtained and combined in order to get the the length of the segment in the 3D space, i.e. the particle range.

\begin{figure}[h!]
\centering
\includegraphics[width=.8\textwidth]{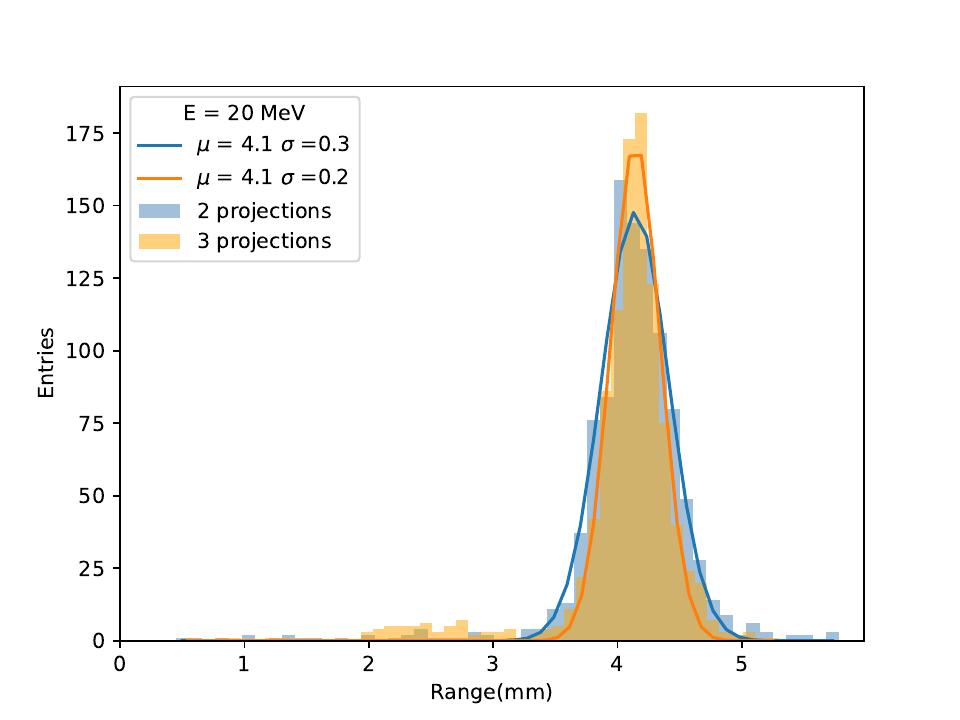}
\qquad
\includegraphics[width=.8\textwidth]     {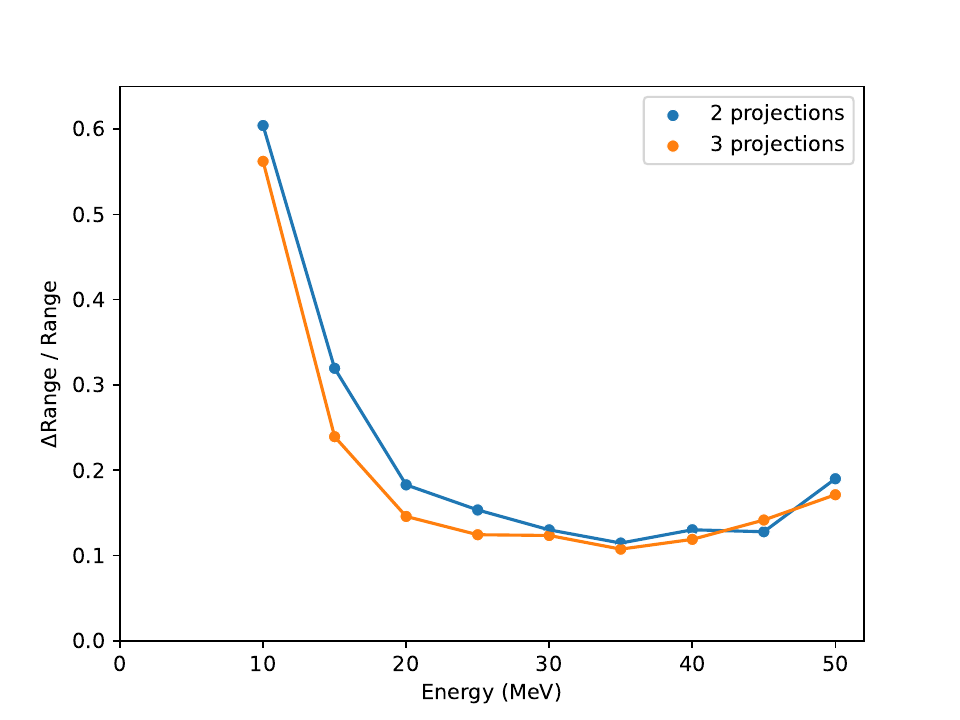}
\caption{Top: example of the range distribution of 20-MeV protons reconstructed with a linear interpolation. Bottom: energy resolution as a function of the proton kinetic energy. 
\label{fig:LI}}
\end{figure}

 In principle, two projections (e.g. on the $x-y$ and $y-z$ planes) are required for a three dimensional reconstruction of a track. However, we expect that the use of a third projection should improve the accuracy of the outcome. 
 Figure~\ref{fig:LI} shows a comparison of track reconstruction using two and three projections. As expected, the energy resolution is slightly better using three projections. 
 For instance, for 20 MeV protons the reconstruction of the particle range improves by a factor 1.5.
  


\subsection{Principal Component Analysis}\label{sec:tracks2}
Principal Component Analysis is a mathematical data analysis technique that is widely used for applications such as dimensionality reduction, lossy data compression, feature extraction and data visualization. It is useful to deal with data sets whose elements can be regarded as points of a certain finite vector space, for current needs considered in the real field. In a broad sense, this tool allows one to find a subspace which is of interest since it is representative of the data set, from a particular point of view.

\begin{figure}[h!]
\centering
\includegraphics[width=0.5\textwidth]{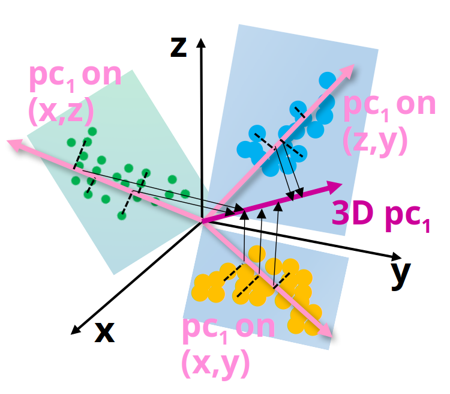}
\caption{Principal components for an idealized 3D data set.}
\end{figure}

PCA can be defined as the orthogonal projection of the data set onto a lower dimensional subspace, whose dimension is imposed according the user need. This principal subspace is obtained such that the variance of the projected data is maximized: namely the projection of the data set is capable of catching its extension. The directions that maximize such variance turn out to be the eigenvectors of the covariance matrix of the data. These eigenvectors can be ordered according to the magnitude of corresponding eigenvalues ~\cite{PCA_book}.

This technique has clear implications in data analysis since one, or several, principal directions can provide insights about the (mutual) relevance of the parameters (coordinates) in the original data vector space.

The main idea behind using PCA for particle track imaging is the projection of the data points on a 1D subspace of the 3D geometrical space, representing in fact the line to which the particle track belongs. Concerning this specific application, PCA has been utilized to produce stereoscopic images of the charged particle tracks, starting from photons collected at the sensor (Sec.~\ref{sec:optic}), taking into account data from 2 or 3 mutually orthogonal planes.

Each 2D projections at the sensor 
is modeled as a bi-dimensional array of squared pixels, counting occurrences of incoming photons. The center of every pixel is a 2D data point with associated weight. Thus, applying PCA at each data set belonging to a
2D projection at the sensor
could yield the principal direction -- mainly the projection of the track on that 2D projection -- and the second orthogonal to the first one. 

\begin{figure}[h]
\centering
\begin{subfigure}{.5\textwidth}
  \centering
  \includegraphics[width=\linewidth]{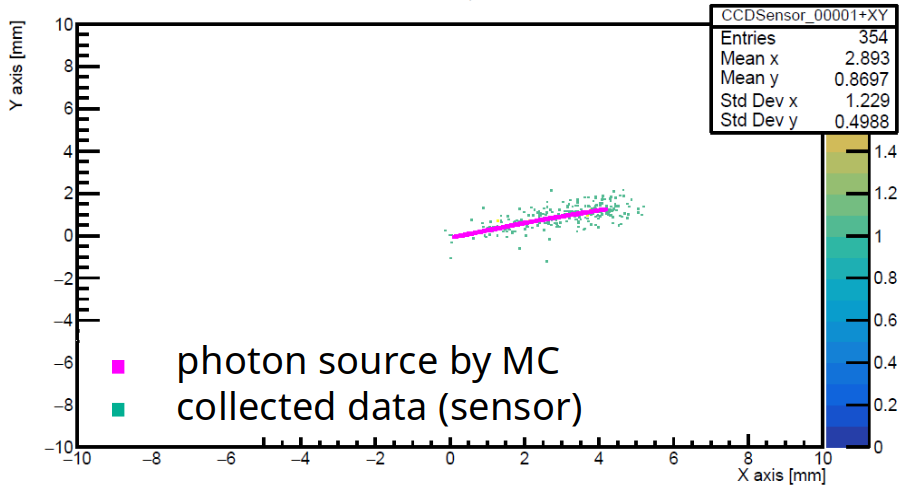}
  \caption{$x-y$ projection at the light sensor.}
\end{subfigure}%
\begin{subfigure}{.5\textwidth}
  \centering
  \includegraphics[width=\linewidth]{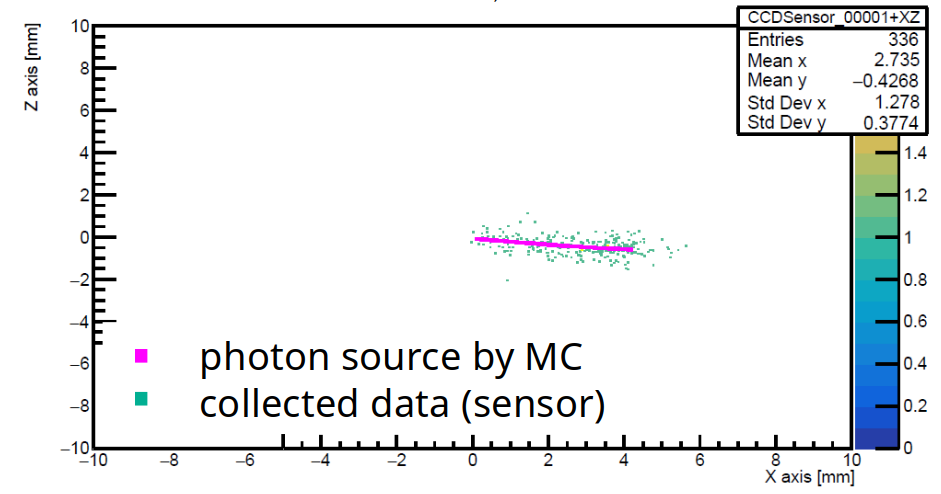}
  \caption{$x-z$ projection at the light sensor.}
\end{subfigure}

\begin{subfigure}{.5\textwidth}
  \centering
  \includegraphics[width=\linewidth]{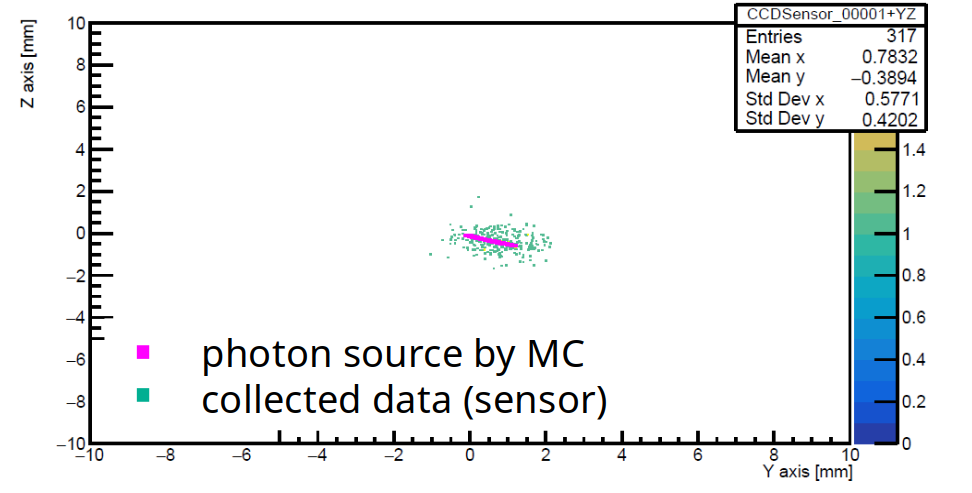}
  \caption{$y-z$ projection at the light sensor.}
\end{subfigure}%
\begin{subfigure}{.5\textwidth}
  \centering
  \includegraphics[width=\linewidth]{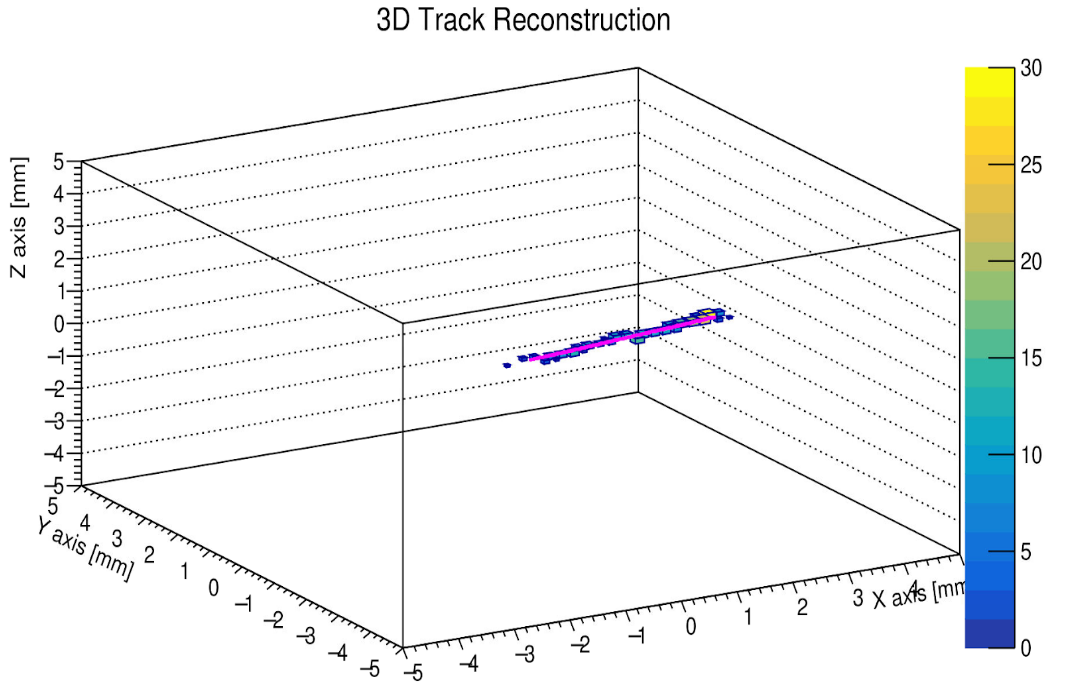}
  \caption{Reconstructed and original (purple) track.}
\end{subfigure}
\caption{Data on three sensor faces, i.e. 2D projections, and reconstructed track of a 30 MeV proton.}
\label{fig:2Dfaces}
\end{figure}

The approach that better takes into account all the available data -- namely from either 2 or 3 projections -- turned out to be the one in which all planes are considered at a time. Each of those produces the covariance between two variables in the 3D space.
The resulting covariant matrix produces the 3D direction of the particle track as the first eigenvector -- namely the principal component. Once the principal direction was obtained in 3D space, it was projected on the sensor plane and then compared with each 2D data set, as showed in fig.~\ref{fig:2Dfaces}. In fig. ~\ref{fig:2Dfaces} data are compared to photon source obtained through Monte Carlo simulations. Every point in the  plane of the sensor is projected on the principal direction projection onto the corresponding face, yielding a point uniquely related to a point on the 3D particle track.

Merging the data sets provided by the faces, the 3D track is reconstructed as well as the distribution of the photon emission events along the particle trajectory as in  the bottom right panel of fig.~\ref{fig:2Dfaces}. Such distribution is in fact related to the ionization capability per unit length and offers a Bragg peak shape from which range particle can be measured and thus the particle energy inferred.

\begin{figure}[h!]
\centering
\includegraphics[width=0.8\textwidth]{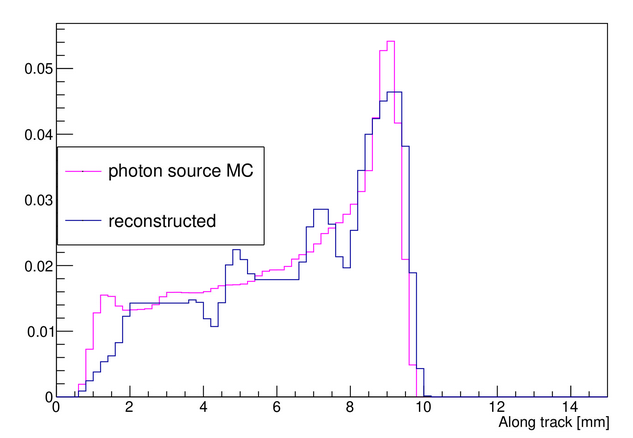}
\caption{Comparison of photon emission along the track of a 30 MeV proton: Geant4 simulation and PCA data analysis. Proton range can be obtained from the distribution and  energy can be inferred.
\label{fig:pca_range}}
\end{figure}

The method showed just before was compared to the vertex at which all photons are produced. An effective check is obtained projecting such photon sources on the particle track previously obtained. Photon sources and processed data are compared, as highlighted in the fig.~\ref{fig:pca_range}

\section{Conclusion}
Riptide is a detector concept conceived for tracking fast neutrons. From two independent analysis of Monte Carlo simulations, we demonstrated the viability of the detector concept based on proton recoil track imaging. The novelty of Riptide with respect to similar detectors in the literature is the use of a  plastic scintillator in combination with CMOS cameras to take a snapshot of the scintillation light. 

The actual number of photons reaching and interacting with the CMOS cameras have to be assessed experimentally in order to confirm its feasibility. Image intensifiers might be used in case of a limited number of counts in the 2D projections.

If successful, Riptide might represent a novel neutron detection technique that can enable a new class of detectors with unprecedented efficiency and timing properties, combined with track reconstruction.

\acknowledgments
The authors acknowledge the use of computational resources from the parallel computing cluster and laboratory infrastructures from the Advanced Sensig Lab of the Open Physics Hub (\url{https://site.unibo.it/openphysicshub/en}) at the Physics and Astronomy Department in Bologna.



\bibliographystyle{JHEP}
\bibliography{Massimi-Riptide_IPRD23.bib}
\end{document}